\documentclass[14pt]{extarticle}

\usepackage[a4paper,margin=2.5cm]{geometry}
\usepackage{amsmath}
\usepackage{amssymb}
\usepackage{graphicx}
\usepackage{siunitx}
\usepackage{bm}
\usepackage{booktabs}
\usepackage{xcolor}
\usepackage{authblk}           
\usepackage{abstract}
\makeatletter
\renewcommand\@maketitle{%
  \newpage\null\vskip 2em%
  \let\footnote\thanks
  {\large\bfseries \@title \par}%
  \vskip 1.5em%
  {\large\raggedright \@author \par}%
  \vskip 1em%
  {\large \@date}%
  \par\vskip 1.5em}
\makeatother
\usepackage[style=numeric,sorting=none,backend=biber]{biblatex}
\usepackage[font=small]{caption}           
\usepackage[colorlinks,citecolor=blue,urlcolor=blue]{hyperref}  

\newcommand{\x}[1]{\mathrm{#1}}

\begin{document}

\title{Graphene-based Photodetector with Engineered Hot Carrier Cooling Dynamics}


\author[1,2]{Yishu Huang}
\author[3]{Anand Nivedan}
\author[3]{Florian Ludwig}
\author[4]{Bohai Liu}
\author[1,2]{Michiel Debaets}
\author[1]{Steven Brems}
\author[5]{Hai I.~Wang}
\author[6]{Alessandro Principi}
\author[1,2]{Dries Van Thourhout}
\author[1]{Christian Haffner}
\author[3]{Aron W.~Cummings}
\author[4,3]{Klaas-Jan Tielrooij\thanks{Corresponding author:
  \href{mailto:k.j.tielrooij@tue.nl}{k.j.tielrooij@tue.nl}}}

\affil[1]{Imec, Kapeldreef 75, 3001 Leuven, Belgium}
\affil[2]{Photonics Research Group, INTEC Department,
  Ghent University--imec, 9052 Ghent, Belgium}
\affil[3]{Catalan Institute of Nanoscience and Nanotechnology (ICN2),
  CSIC and BIST, Campus UAB, Bellaterra, Barcelona 08193, Spain}
\affil[4]{Department of Applied Physics, TU Eindhoven,
  Den Dolech 2, 5612 AZ Eindhoven, The Netherlands}
\affil[5]{Debye Institute for Nanomaterials Science,
  Utrecht University, 3584 CS Utrecht, The Netherlands}
\affil[6]{Department of Physics and Astronomy, The University of Manchester,
  Oxford Road, Manchester M13 9PL, United Kingdom}
\date{}

\maketitle

\begin{abstract}\normalsize
Graphene has emerged as a promising material for integration into silicon photonics,
owing to its ultrafast and broadband photoresponse without the need for an external
bias voltage.
This photoresponse relies on the photo-thermoelectric effect created by hot carriers.
A key factor underlying the performance of graphene photodetectors is the cooling
dynamics of these hot carriers.
In this work, we engineer these dynamics in a WSe$_2$-graphene-WSe$_2$
waveguide-integrated photodetector.
In particular, by introducing proximity screening by a nearby graphite layer to this
structure, we prolong the hot-carrier cooling time, leading to an enhanced
photoresponse.
We characterize the cooling dynamics under continuous-wave laser excitation by
employing a photomixing technique, revealing an increase in the cooling time by up
to a factor of four.
Direct photoresponse measurements show that the internal photoresponsivity improves
by approximately 50\%.
Together, these results demonstrate the potential of proximity screening to enhance
the performance of graphene-based photodetectors on an integrated photonics platform.
\end{abstract}

\bigskip

\begin{refsection}  

\section{Introduction}

Photodetectors facilitate the conversion of optical signals into electrical signals,
a capability widely required in both long-haul optical communication systems and
short-range data center applications.
Graphene is a promising material for photodetection owing to its zero bandgap, high
carrier mobility, low electronic heat capacity~\cite{Akinwande2019,Aamir2021,
Pospischil2013}, and its ability to generate a direct electron-heat-driven
photovoltage signal without requiring an external bias~\cite{doi:10.1126/science.1211384,
Song2011}.
This strongly reduces the dark current and removes the need for a transimpedance
amplifier~\cite{Marconi2021,Schuler2021,Yu2023,Muench2019}.
Graphene-based photodetectors also offer extremely fast operation, up to hundreds of
GHz~\cite{Koepfli2024}.
However, in order to compete with commercial photodetectors, such as those based on
germanium, it is necessary for graphene-based photodetectors to increase their
photoresponsivity~\cite{Lischke2021}.

To enhance the photoresponsivity of graphene photodetectors, several approaches have
been explored, mostly aimed at increasing light absorption in graphene, including
optical ring resonators~\cite{Schuler2021} and plasmonic nanostructures~\cite{Muench2019}.
Alternatively, it has recently been proposed that slower cooling of hot carriers,
tailored via proximity screening, can lead to an increased photoresponse, as it leads
to an increased average carrier temperature~\cite{Wang2026}.
The resulting steeper temperature gradients enhance the thermally driven transport of
photo-generated carriers, thereby strengthening the photoresponse.
This strategy of increasing the photoresponse through slower hot carrier cooling
requires control over the cooling dynamics, which have been reviewed in
Ref.~\cite{Massicotte2021}.

Recent work demonstrated that high-quality WSe$_2$-encapsulated graphene exhibits
longer cooling times than hexagonal boron nitride (hBN)-encapsulated graphene, with
the cooling dynamics ultimately limited by the emission of optical
phonons~\cite{doi:10.1021/acsnano.0c10864}.
In this scenario, rapid carrier--carrier scattering efficiently re-thermalizes the
electronic system, maintaining relatively fast optical-phonon-mediated cooling that
occurs on a picosecond timescale~\cite{doi:10.1021/acsnano.0c10864}.
Motivated by very recent work demonstrating the slowing of cooling dynamics in a
high-permittivity environment~\cite{Wang2026}, enhanced responsivities in
photodetectors based on WSe$_2$-encapsulated graphene~\cite{SamyArxiv2026}, and
enhanced quantum mobilities due to proximity screening~\cite{Domaretskiy2025},
further improvements to the photoresponse seem to be attainable.

These encouraging developments make an experimental demonstration of proximity
screening effects in graphene photodetectors integrated on a silicon photonic platform
highly desirable, in order to demonstrate relevance in real-world applications.
However, until now, this has been lacking.
In this work, we investigate the influence of proximity screening from a nearby
graphite layer on a WSe$_2$--graphene--WSe$_2$ waveguide-integrated photodetector.
We experimentally compare the cooling dynamics and photoresponse of the exact same
device before and after placing the graphite layer, corresponding to the situation
without and with proximity screening, respectively.
Since the photodetectors of datacommunication and telecommunication applications
operate under low-power illumination, we characterize the cooling dynamics using a
continuous-wave (CW)-laser--based photomixing technique, based on
Ref.~\cite{Jadidi_PRL}, rather than measurements with ultrashort pulses, as these
correspond to high peak powers~\cite{Massicotte2021}.
Our results show that the introduction of proximity screening leads to an increase
in the cooling time by a factor of 2--4, and a 1.5-fold enhancement in the graphene
photoresponsivity.
This shows that engineering the hot carrier cooling dynamics is indeed a promising
strategy for enhancing the performance of graphene-based, silicon-photonics-integrated
photodetectors.

\section{Results}

\subsection{\texorpdfstring{WSe$_2$-graphene-WSe$_2$}{WSe2-graphene-WSe2}
  waveguide photodetector}

\begin{figure*}[htbp]
  \centering
  \includegraphics[width=\textwidth]{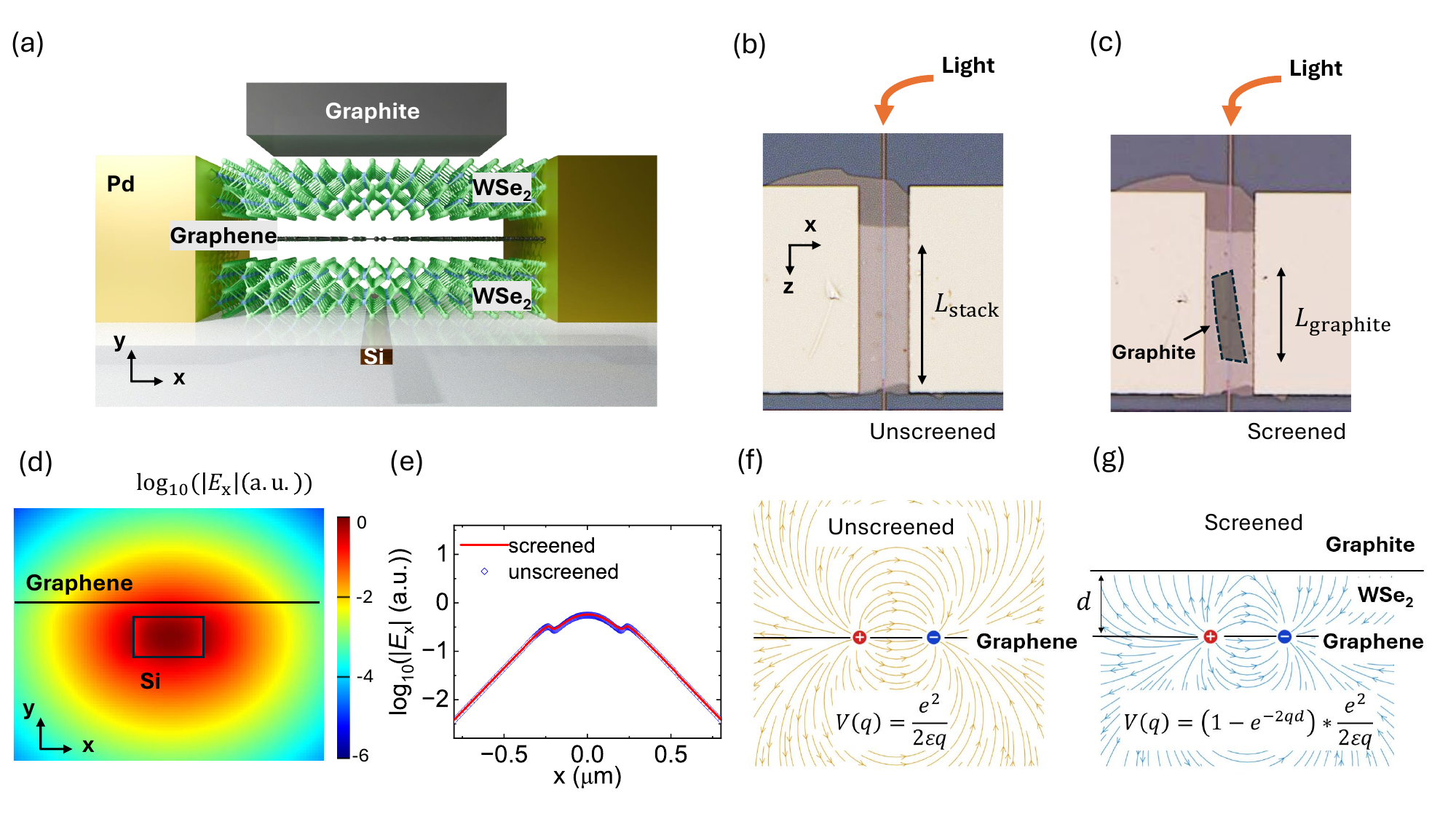}
  \caption{%
    (a)~Schematic of the cross section of the measured device, consisting of
    WSe$_2$-encapsulated graphene placed on a silicon waveguide, and a graphite
    screening layer on top.
    (b--c)~Top view of the WSe$_2$--graphene--WSe$_2$ waveguide photodetector
    for the unscreened~(b) and screened~(c) cases.
    (d)~Cross section of the optical field profile in the waveguide.
    (e)~The optical field in graphene for the unscreened and screened cases.
    (f--g)~Illustration of the physical mechanism behind the proximity screening
    effect, for the example of an electron and a hole, for the unscreened~(f) and
    screened~(g) case.
    The field lines are affected by the graphite layer, suppressing the strength of
    the carrier--carrier interaction.
    The equations describe the Coulomb interaction among carriers in the (un)screened
    case, where $q$ is the carrier momentum, $e$ denotes the elementary charge,
    $\epsilon$ represents the permittivity, and $d$ is the distance between graphene
    and graphite.}
  \label{fig:1}
\end{figure*}

Figure~\ref{fig:1}a shows the schematic of our device, a
WSe$_2$-graphene-WSe$_2$ waveguide-integrated photodetector with graphite
screening layer on top.
We prepared a stack of WSe$_2$-graphene-WSe$_2$, which we dry-transferred to a
silicon optical waveguide designed for operation at the telecom wavelength
$1550\,\mathrm{nm}$, forming the channel.
The silicon optical waveguide has dimensions of
$450\,\mathrm{nm}\times 220\,\mathrm{nm}$ and is covered by a few-nanometer-thick
planarized oxide top cladding.
We use WSe$_2$ as the encapsulating material, rather than hBN, because it
suppresses hyperbolic phonon mediated cooling, making intrinsic optical phonon
cooling the dominant relaxation
channel~\cite{doi:10.1021/acsnano.0c10864,Principi2017,Tielrooij2018}.
The graphene is exfoliated from high-quality crystals, to minimize disorder and
defects, thereby suppressing supercollision
cooling~\cite{Song_2012_PRL,Betz2013}.
The dry-transfer process~\cite{Kinoshita2019}, assisted by polycarbonate/PDMS
stamps, sequentially picked up the flakes to assemble the stack, which was then
transferred onto the silicon waveguide.
Following the dry-transfer process, plasma etching and metal deposition were
performed to form edge contacts~\cite{edge_contact_science.1244358} to the
WSe$_2$-encapsulated graphene, completing the device for the unscreened case.
For the screened case, a graphite layer was deposited on top through an additional
dry-transfer step.

Figure~\ref{fig:1}b shows the top view of the device in the unscreened case,
where the stack length is $L_{\mathrm{stack}} = 40\,\mu\mathrm{m}$.
A longer stack leads to increased absorption, yet not necessarily a stronger
photoresponse.
Figure~\ref{fig:1}c presents the top view of the screened configuration, in which
the graphite screening layer, which has a length of
$L_{\mathrm{graphite}} = 25\,\mu\mathrm{m}$, is placed on top.
Light propagates along the $z$-direction within the silicon waveguide.
As it passes through the stacked region, the evanescent field overlaps with the
graphene layer and is partially absorbed, generating a photoresponse.
The cross section of the optical field profile in the waveguide is shown in
Fig.~\ref{fig:1}d, while Fig.~\ref{fig:1}e shows the optical field in the graphene
channel, highlighting a minimal difference between the unscreened and screened cases.
The simulated optical mode profile changes only weakly due to the graphite,
indicating that the main role of the graphite layer is proximity screening.

The concept of proximity screening is illustrated in Fig.~\ref{fig:1}(f--g), for
the example of an electron and a hole, where the field lines---including those in
between the electron and hole---are affected by the presence of the graphite layer.
The Coulomb interaction among carriers is given by $V(q) = e^2/(2\epsilon q)$
(unscreened) and
$V(q) = (1-\x{e}^{-2qd})\cdot e^2/(2\epsilon q)$ (screened), where $q$ is the
carrier momentum, $e$ denotes the elementary charge, $\epsilon$ represents the
permittivity, and $d$ is the distance between graphene and graphite.
The presence of nearby graphite imposes a small $d$, which suppresses the Coulomb
interaction via proximity screening.
This screening weakens carrier--carrier scattering in graphene, which in turn slows
down the re-thermalization processes associated with optical phonon cooling.
As a result, the cooling time of photo-generated hot carriers is prolonged, as shown
in Ref.~\cite{Wang2026}.
Under steady-state optical excitation, this leads to a higher carrier temperature,
enhancing the electron-heat-driven photovoltages.

To verify this proximity screening effect on the cooling dynamics and the
photoresponse, we characterized the cooling time on the same device for both the
unscreened and screened cases using a CW laser technique, thereby closely mimicking
realistic datacommunication and telecommunication operating conditions.
We also performed photoresponse measurements for both the unscreened and screened
cases to evaluate if the prolonged cooling time leads to enhanced performance in terms
of the responsivity.

\subsection{Cooling time characterization}

\begin{figure*}[htbp]
  \centering
  \includegraphics[width=\textwidth]{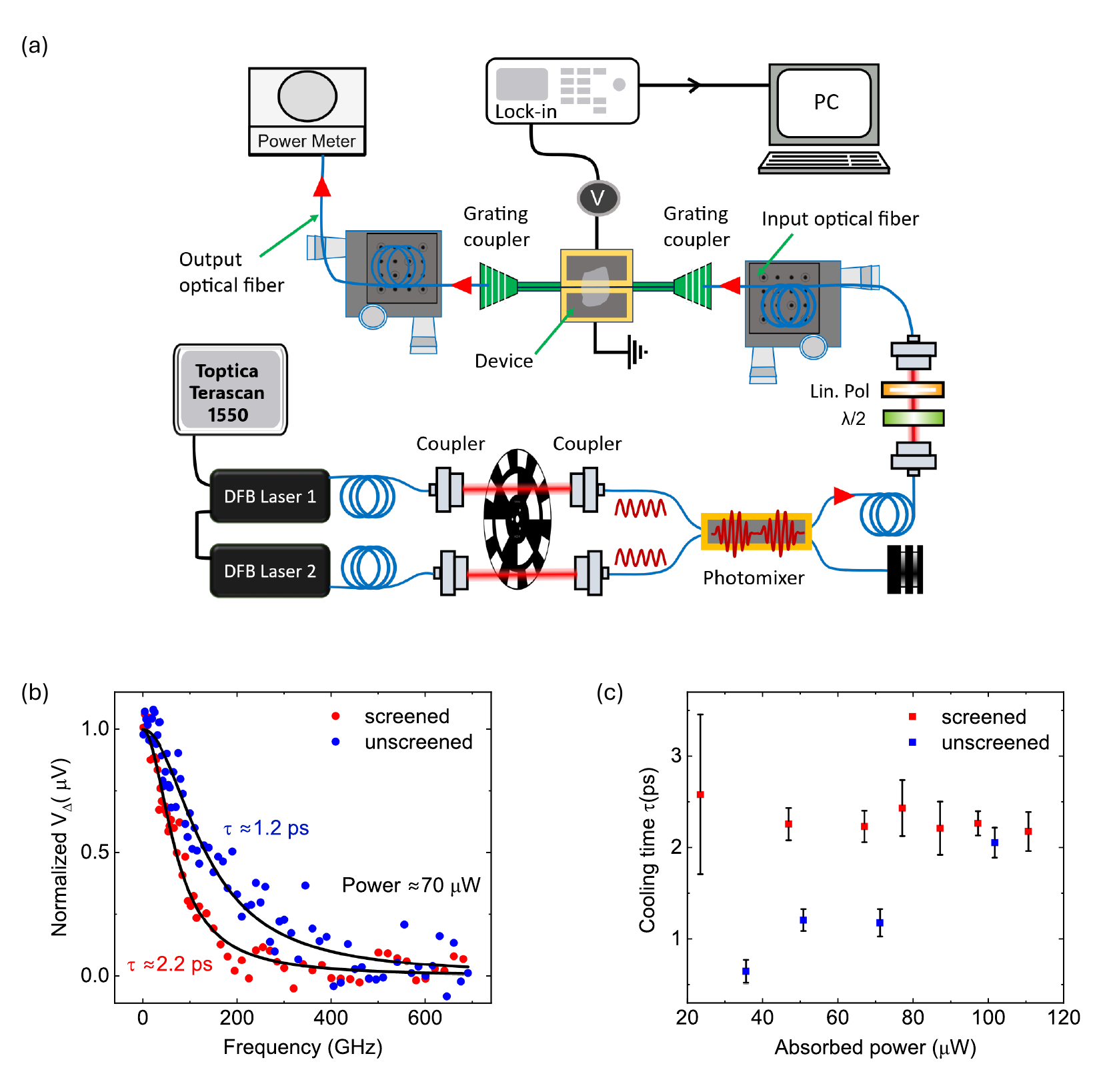}
  \caption{%
    (a)~Schematic of the experimental setup.
    Two continuous-wave distributed feedback lasers with a tunable difference
    frequency are combined and coupled into the waveguide of the device via grating
    couplers.
    The incident light is mechanically modulated with a chopper, and the resulting
    mixing photovoltage ($V_{\Delta}$) is detected using a lock-in amplifier.
    (b)~Normalized, background-subtracted photovoltage response as a function of the
    heterodyne beat frequency.
    The data points represent the measured response for the device in the unscreened
    (blue circles) and screened (red circles) cases.
    The solid lines indicate Lorentzian descriptions used to extract the hot-carrier
    cooling time $\tau$, yielding $\tau \approx 1.2\,\mathrm{ps}$ and
    $\tau \approx 2.2\,\mathrm{ps}$ for the unscreened and screened cases,
    respectively, at an absorbed power of $\approx 70\,\mu\mathrm{W}$.
    (c)~Measured hot-carrier cooling time $\tau$ as a function of absorbed power for
    the device in the unscreened (blue squares) and screened (red squares) cases.
    The screened device shows a longer and more stable cooling time over the measured
    power range.
    The error bars are based on multiple measurements as a function of difference
    frequency and represent the 68\% confidence interval.}
  \label{fig:2}
\end{figure*}

We study the hot-carrier cooling dynamics using heterodyne photomixing
(see Fig.~\ref{fig:2}a).
Two continuous-wave distributed feedback (DFB) lasers with a tunable frequency mix
to form a beat note, which is then coupled into the optical waveguide of the graphene
device through a grating coupler.
The optical beat note periodically modulates the absorbed power, causing the carrier
temperature to oscillate at the beat frequency, which is given by the relative
frequency detuning of the two DFB lasers.
Because the photovoltage depends nonlinearly on the carrier temperature, this produces
a time-averaged beating photovoltage, which we measure using a lock-in amplifier.
At low beat frequencies, the carriers can follow the modulation, whereas at high beat
frequencies they cannot respond fast enough, causing the mixing signal to roll off.
The cooling time is extracted from this frequency-dependent roll-off~\cite{Jadidi_PRL}.
This continuous-wave method probes the device in a low-$\Delta T$ regime, where the
electron temperature does not increase much beyond the lattice temperature, closer to
realistic device operating conditions.
It therefore provides direct access to the intrinsic carrier cooling dynamics without
strongly perturbing the electron distribution.

Figure~\ref{fig:2}b shows the normalized heterodyne photovoltage response as a
function of beat frequency for the device with and without graphite screening.
The solid lines are descriptions with a Lorentzian function of the frequency-domain
response of the beating photovoltage, where the width is inversely proportional to
the cooling time~\cite{Jadidi_PRL} (see also Supplementary Note~2).
The data shown in Fig.~\ref{fig:2}b correspond to the same absorbed power of
$\approx 70\,\mu\mathrm{W}$ in both the unscreened and screened cases.
At this power, the extracted cooling time increases from $\tau \approx 1.2\,\mathrm{ps}$
in the unscreened case to $\tau \approx 2.2\,\mathrm{ps}$ in the screened case.
In Supplementary Note~4, we show a detailed simulation of the full experiment,
relating the obtained cooling time from the frequency roll-off to a microscopic
cooling time that serves as input to the simulation of the experiment.
We find that the photomixing technique tends to somewhat underestimate the cooling
time, in contrast to time-resolved photocurrent measurements, which tend to somewhat
overestimate the cooling time.
This means that the cooling times that we obtain here are lower bounds.

Figure~\ref{fig:2}c summarizes the extracted cooling times as a function of the
absorbed optical power in graphene.
The absorbed power is quantified by accounting for the excess waveguide propagation
losses (see Supplementary Note~3).
These losses are higher in the screened case than in the unscreened case, which is
due to optical absorption in the graphite screening layer.
The screened device exhibits a longer cooling time than the unscreened device over
the measured power range, and its cooling time remains nearly independent of power.
This contrasts with the clear power dependence of $\tau$ observed in the unscreened
case.
We attribute the longer cooling time in the screened case to the proximity screening
of carrier--carrier interactions, as proposed in Ref.~\cite{Wang2026}.
This weakens carrier--carrier scattering and reduces the efficiency of electronic
re-thermalization within the optical-phonon-mediated cooling dynamics, thereby
increasing the effective hot-carrier cooling time.
In contrast, in the unscreened case, electronic re-thermalization remains efficient,
enabling optical-phonon emission and the onset of a hot-phonon bottleneck, as shown
in Ref.~\cite{doi:10.1021/acsnano.0c10864}, which leads to a stronger power
dependence of the cooling time.

\subsection{Photoresponse}

\begin{figure*}[htbp]
  \centering
  \includegraphics[width=\textwidth]{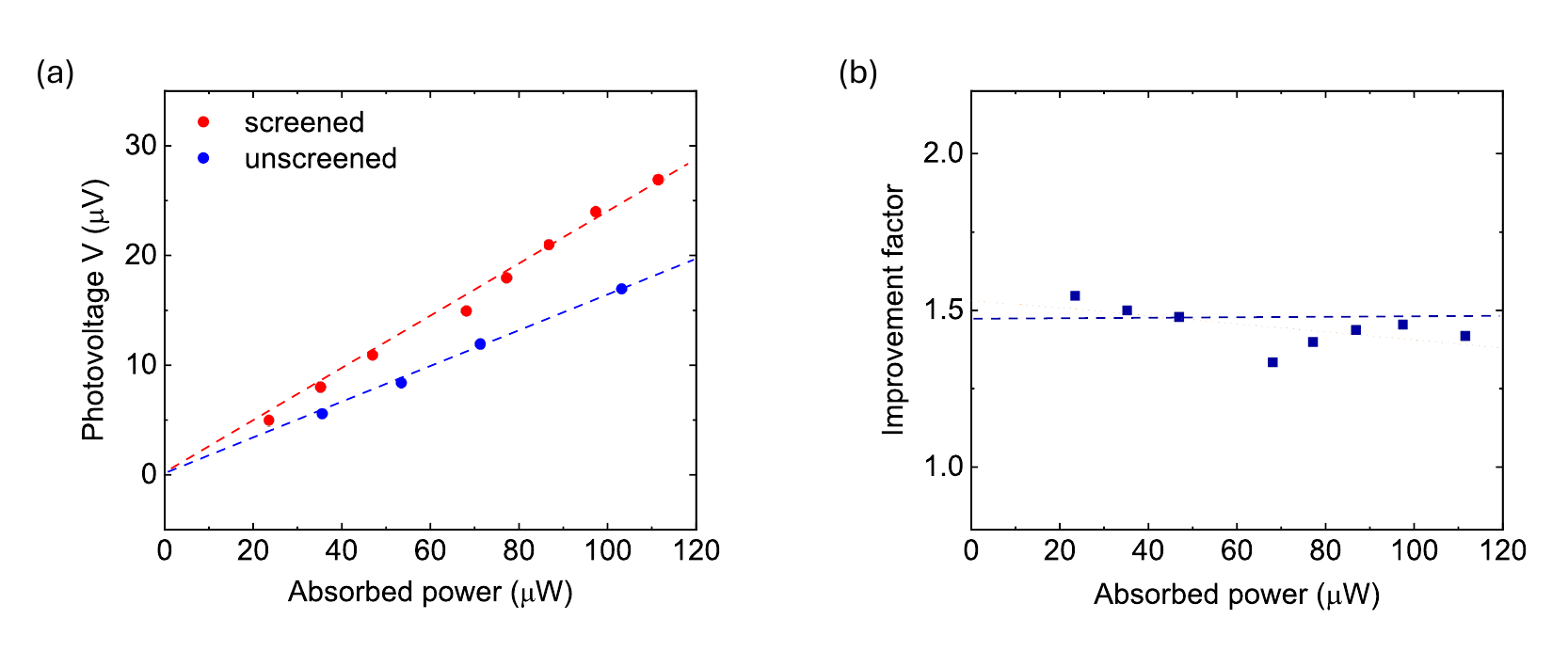}
  \caption{%
    (a)~Measured photovoltage as a function of absorbed power for the device in the
    unscreened (blue circles) and screened (red circles) cases, with dashed lines
    indicating linear fits.
    (b)~Improvement factor as a function of absorbed power, with the dashed line
    indicating the average value of $\approx 1.5$.}
  \label{fig:3}
\end{figure*}

With the cooling time shown to increase after adding a graphite screening layer,
we next compare the photoresponse of the device before and after adding the screening
layer.
As shown in Fig.~\ref{fig:3}a, the device with the graphite screening layer exhibits
a higher photovoltage for a given absorbed power, demonstrating a clear improvement
of internal photoresponsivity.
In both cases, the photovoltage increases monotonically with absorbed power, with no
clear saturation over the measured range.

Figure~\ref{fig:3}b shows the measured improvement factor, which is approximately
1.5.
This improvement factor can be explained by using the measured cooling-time data as
input for simulations based on the photothermoelectric effect driven by hot carriers
in graphene~\cite{Song2011,Massicotte2021,Antidormi2021} (see Supplementary Note~4).
It confirms that the measured cooling time data and the measured photoresponse are
consistent with our hypothesis that proximity screening suppresses the efficiency of
cooling channels, effectively ``trapping'' heat within the electronic system.

\section{Discussion}

Having demonstrated slower cooling and an increased photoresponse enabled by screening
of carrier--carrier interactions by a nearby graphite layer, we now discuss strategies
to further improve device performance by exploiting the proximity screening effect.
One approach to increase the screening of carrier--carrier scattering is to place the
graphite screening layer very close to the graphene channel, on the order of
${\sim}1\,\mathrm{nm}$.
Such extremely close proximity screening configurations have been studied in studies
aimed at (hydrodynamic) charge transport~\cite{Domaretskiy2025,Kim2020}; however,
it is not yet clear to what extent this enhanced screening would influence the carrier
cooling time and the photoresponse.

To estimate this, we refer to very recent work that examined the impact of dielectric
screening on cooling times in graphene~\cite{Wang2026}.
For pure dielectric screening, numerical simulations of the cooling dynamics reveal
that the cooling time scales approximately linearly with permittivity,
$\tau \propto \epsilon$ (see Supplementary Note~5).
In the presence of proximity screening and for small graphene--graphite distance $d$,
the Coulomb interaction $V(q) \approx e^2 d/\epsilon$, such that the cooling time
scales as $\tau \propto 1/d$.

In the limit of a monolayer WSe$_2$ spacer, the distance between graphene and
graphite is $d \approx 1.3\,\mathrm{nm}$, determined by the monolayer WSe$_2$
thickness and the van der Waals spacing.
This means that we could expect an additional $4.7\times$ enhancement of the cooling
time beyond our measured results for which $d \approx 6.1\,\mathrm{nm}$, yielding
approximately a $12\times$ enhancement of cooling time compared to the unscreened
case.
If we assume a photodetector operating under the photo-thermoelectric effect in
graphene, where the photovoltage scales as the square root of the cooling
time~\cite{Song2011,Wang2026,Antidormi2021}, this would correspond to a
$3.5\times$ enhancement of the photoresponse.

\section{Conclusions}

In this work, we demonstrate that proximity screening can prolong the cooling time of
hot carriers in graphene, leading to improved internal photoresponsivity.
Specifically, by introducing a graphite screening layer at a distance of
$6.1\,\mathrm{nm}$ from graphene, the carrier cooling time is prolonged by up to a
factor of four at low optical powers, resulting in a ${\sim}50\%$ enhancement of the
intrinsic photoresponse.
Further improvement in responsivity with an extremely nearby proximity layer is
expected, which is possibly enough to outcompete several competing receiver technology
platforms.
In addition to exploring extremely proximate screening layers, future work should
address the challenge of maintaining optimal absorption in the graphene channel.
The fact that our device was fabricated on a standard waveguide platform and
characterized using a CW-laser-based photomixing technique highlights the potential
for datacommunication and telecommunication applications.

\section*{Acknowledgments}
This work was funded by FLAG-ERA project ``ENPHOCAL'', by MICIN with
No.~PCI2021-122101-2A (Spain) and FWO SBO with No.~S008421N (Belgium).
The ICN2 is supported by the Severo Ochoa Centres of Excellence programme,
Grant CEX2021-001214-S, funded by MCIU/AEI/10.13039.501100011033.
K.-J.T.\ acknowledges funding from the European Union's Horizon~2020 research and
innovation program under Grant Agreement No.~101125457 (ERC CoG ``EQUATE'').

\section*{Author contributions}
Y.H.\ designed and fabricated the device.
A.N.\ performed the measurements.
Y.H., A.N., and K.-J.T.\ interpreted the results, with input from D.V.T.\ and C.H.
A.W.C.\ and A.P.\ performed the simulations of the photodetector response and cooling
dynamics, with input from K.-J.T.
F.L.\ performed the simulations of the photomixing experiment.
B.L., S.B., H.I.W., and M.D.\ provided input for the fabrication and device aspects.
Y.H., A.N., A.W.C., D.V.T., and K.-J.T.\ wrote the manuscript, with input from all
authors.
D.V.T., C.H., and K.-J.T.\ supervised the work.
K.-J.T.\ conceived the idea and coordinated the project.

\section*{Data availability}
All data used to generate the figures in this study are available from the
corresponding author on reasonable request.

\printbibliography
\end{refsection}

\clearpage

\setcounter{equation}{0}
\setcounter{figure}{0}
\setcounter{table}{0}
\renewcommand{\theequation}{S\arabic{equation}}
\renewcommand{\thefigure}{S\arabic{figure}}
\renewcommand{\thetable}{S\arabic{table}}

\begin{center}
  {\LARGE\textbf{Supplementary Information}} \\[0.5em]
  
\end{center}

\bigskip

\begin{refsection}  

\section*{Note 1. Device Fabrication}

\subsection*{AFM data of flakes}

Monolayer graphene and few-layer WSe$_2$ flakes (top and bottom layers) were mechanically exfoliated from bulk crystals.
Optical microscope images of these flakes are presented in Fig.~S1a--c, corresponding to monolayer graphene, the top WSe$_2$ few-layer flake, and the bottom WSe$_2$ few-layer flake, respectively.
Atomic force microscopy (AFM) measurements of the WSe$_2$ layers are shown in Fig.~S1d and Fig.~S1e for the top and bottom layers, respectively.
The extracted thicknesses are approximately $5.5\,\mathrm{nm}$ for the top layer and $5.0\,\mathrm{nm}$ for the bottom layer.
The AFM height measurement of graphene is not included, as the standard tapping mode used in our AFM setup does not provide reliable thickness determination for monolayer graphene.

\begin{figure}[htbp]
  \centering
  \includegraphics[width=\textwidth]{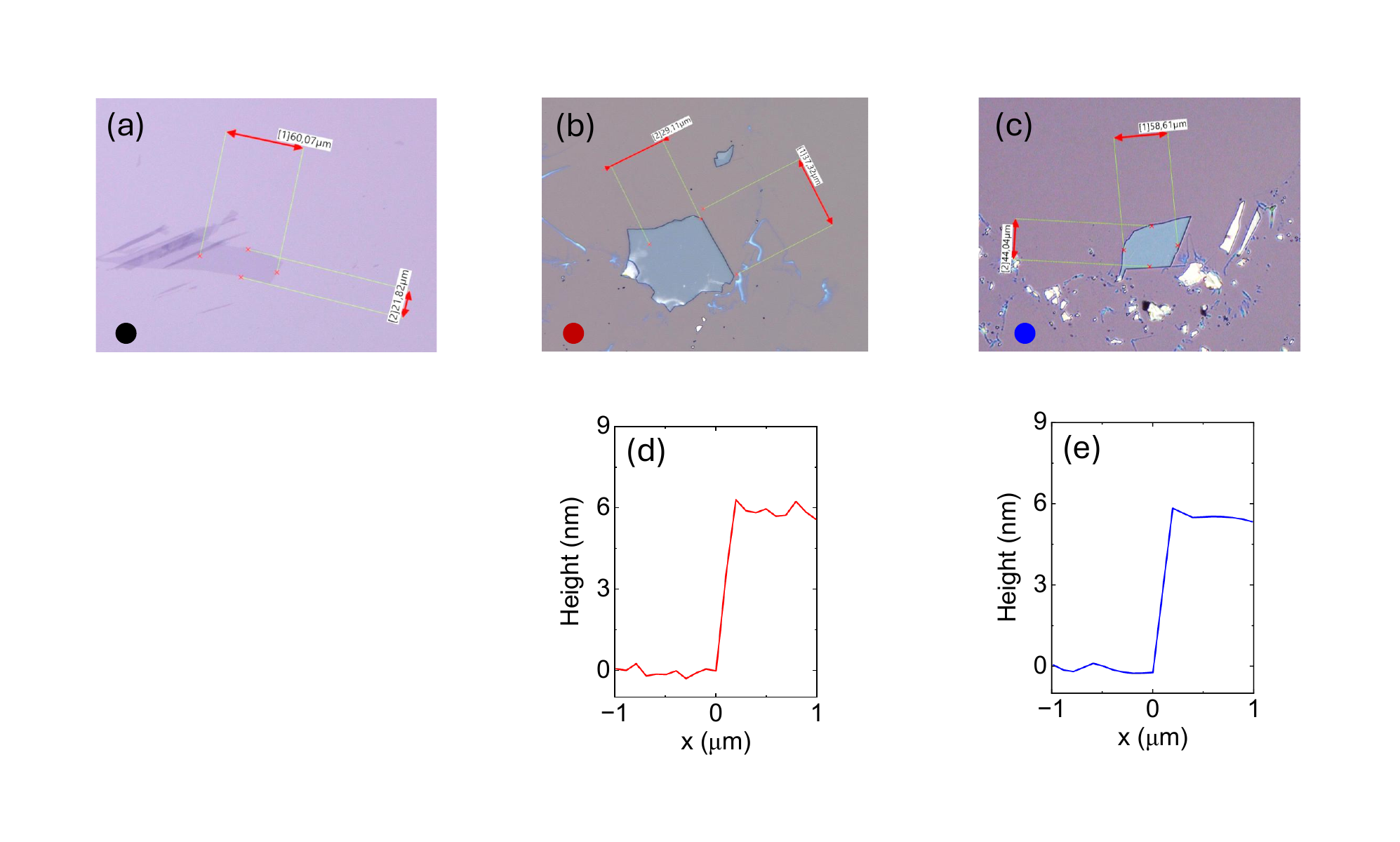}
  \caption{Optical microscope images of (a)~monolayer graphene, (b)~top WSe$_2$ few-layer flake,
           and (c)~bottom WSe$_2$ few-layer flake.
           (d,~e)~Atomic force microscopy (AFM) height profiles of the top and bottom
           WSe$_2$ layers, respectively.}
  \label{sfig:1}
\end{figure}

\subsection*{Raman of \texorpdfstring{WSe$_2$}{WSe2}-graphene-\texorpdfstring{WSe$_2$}{WSe2}}

To verify the monolayer nature and quality of WSe$_2$-encapsulated graphene, a Raman measurement was conducted.
The Raman check point and the corresponding Raman spectrum are shown in Fig.~S2a and Fig.~S2b, respectively.
The full width at half maximum (FWHM) of the 2D peak is $20\,\mathrm{cm}^{-1}$, indicating low disorder in the encapsulated graphene and---as a result---suppressed supercollision cooling~\cite{Song_2012_PRL}.

\begin{figure}[htbp]
  \centering
  \includegraphics[width=\textwidth]{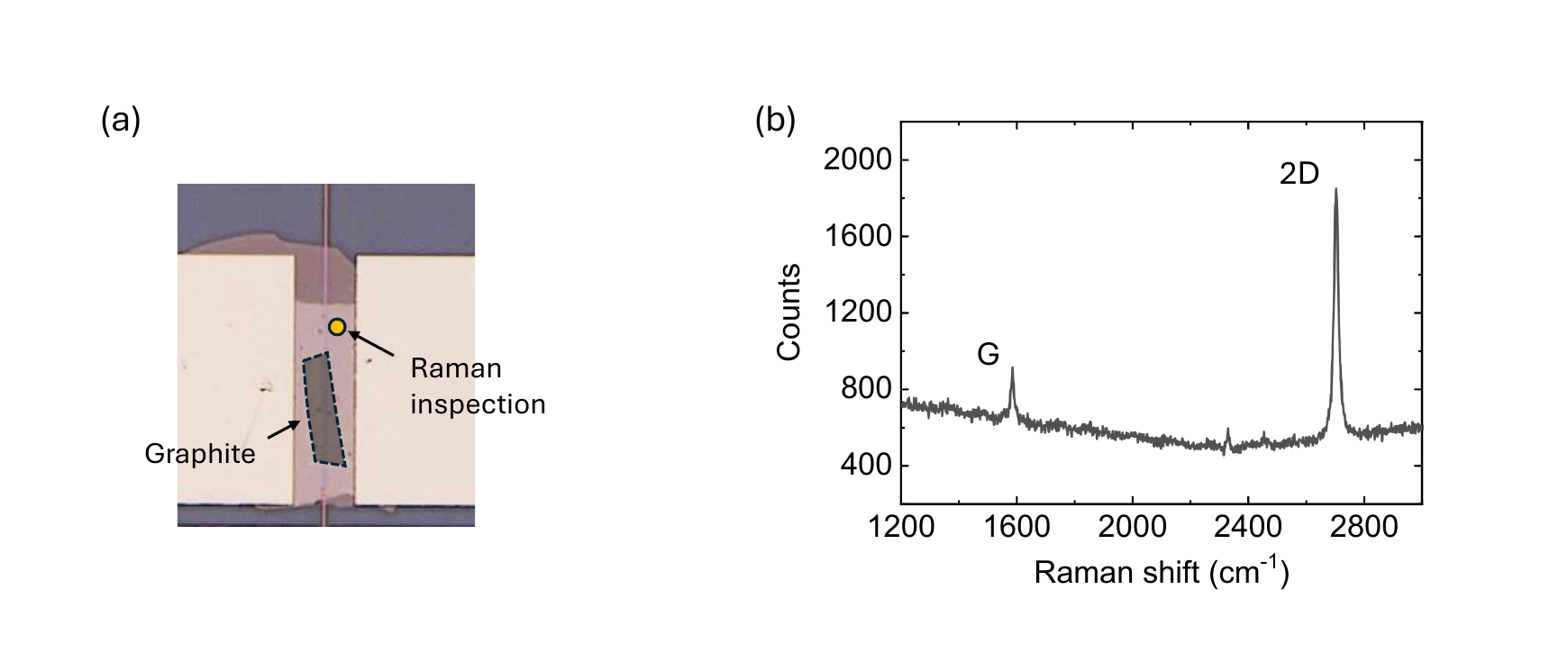}
  \caption{(a)~Microscope image and (b)~Raman spectrum of the spare
           WSe$_2$-graphene-WSe$_2$ stack.}
  \label{sfig:2}
\end{figure}

\section*{Note 2. Photomixing measurement}

Figure~S3 shows the heterodyne photomixing measurements of the normalized photovoltage
$V_{\Delta}$ as a function of beat frequency for both the unscreened (Fig.~S3a) and
proximity-screened (Fig.~S3b) devices, recorded over a range of input optical powers.
The solid symbols represent the measured data, while the solid lines are fits used to
extract the hot-carrier cooling time $\tau$.

The observed frequency dependence arises from the finite cooling time of hot carriers.
Within the thermal model, the response follows a Lorentzian form in angular
frequency~\cite{Jadidi_PRL},
\begin{equation}
    \frac{\gamma^2}{\omega^2 + \gamma^2},
\end{equation}
where $\omega = \omega_1 - \omega_2$ is the optical beat frequency and
$\gamma = 1/\tau$ is the carrier cooling rate.
This explicitly shows that the frequency-dependent photovoltage is governed by a
Lorentzian response, with the width set by the cooling dynamics.

Experimentally, the data are analyzed as a function of the linear frequency
$f = \omega/2\pi$.
In this representation, the photovoltage is described by
\begin{equation}
    V_{\Delta}(f) = V + \frac{A}{1 + (2\pi f \tau)^2},
\end{equation}
where $V$ is the high-frequency asymptotic value and $A$ is the amplitude of the
frequency-dependent contribution.
Since only positive beat frequencies ($f \ge 0$) are experimentally accessible,
the measured response corresponds to one half of a symmetric Lorentzian profile.

\begin{figure}[htbp]
  \centering
  \includegraphics[width=0.75\textwidth]{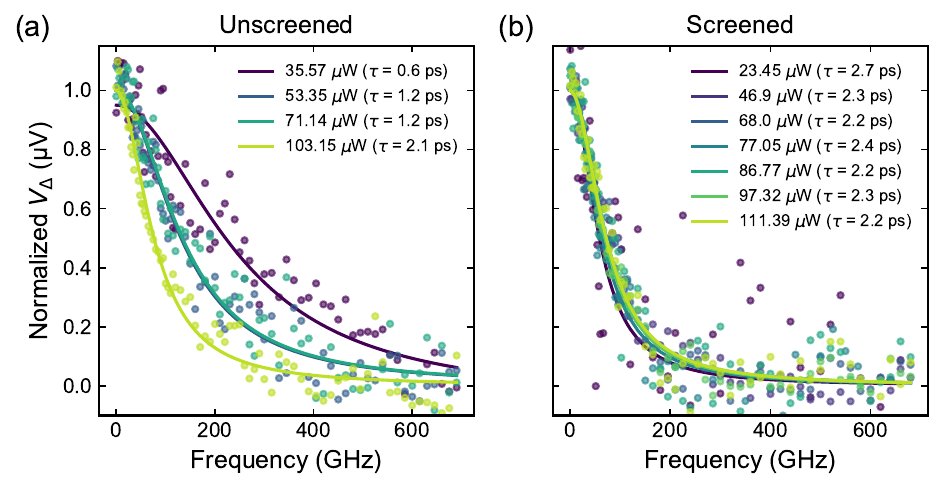}
  \caption{Normalized photovoltage $V_{\Delta}$ as a function of the heterodyne beat
           frequency for the (a)~unscreened and (b)~screened photodetector device.
           Data are plotted for various absorbed optical powers.
           Solid lines represent Lorentzian fits used to extract the hot-carrier
           cooling time $\tau$.}
  \label{sfig:3}
\end{figure}

The characteristic roll-off frequency in Fig.~S3 reflects the inverse of the cooling
time.
In the unscreened device, the roll-off shifts significantly towards lower frequencies
with increasing optical power, indicating a strong increase in $\tau$.
In contrast, the screened device (Fig.~S3b) exhibits only a weak shift of the
roll-off frequency, demonstrating that the cooling time remains nearly constant over
the same power range.
This difference directly highlights the suppression of power-dependent cooling
dynamics in the presence of screening.

The heterodyne excitation is generated by two continuous-wave DFB lasers with angular
frequencies $\omega_1$ and $\omega_2$, producing a beat signal at
$\omega = \omega_1 - \omega_2$.
The resulting absorbed power is given by
\begin{equation}
    P(t) = P_1 + P_2 + 2\sqrt{P_1 P_2}\cos(\omega t),
\end{equation}
where $P_1$ and $P_2$ are the individual laser powers.
Due to the nonlinear dependence of the photothermoelectric voltage on electron
temperature, this oscillating power leads to a rectified DC photovoltage.
The finite carrier cooling time introduces a delayed response to this modulation,
resulting in a suppression of the photovoltage at higher beat frequencies.
This suppression follows the Lorentzian dependence described above, allowing direct
extraction of $\tau$ from the frequency-dependent measurements.

\section*{Note 3. Graphene absorbed power}

To quantify the optical power absorbed by graphene, the excess propagation loss from
the stack and the graphite screening layer must be characterised.
The procedure is described below.

\textit{Stack loss.}
The measured propagation loss for the unscreened case is $0.375\,\mathrm{dB/\mu m}$,
consisting of contributions from graphene absorption and excess loss.
The excess loss arises from polymer residues introduced during the dry-transfer
process and from optical scattering within the stack.
The stack exhibits wrinkles, likely induced by lattice mismatch between WSe$_2$ and
graphene~\cite{WSe2_Gra_bandalinment}, which contribute to the observed scattering
loss.

\textit{Graphite loss.}
For the screened case, the additional optical loss of the graphite layer reduces the
output power below the detection limit of our power meter, necessitating a separate
loss characterisation.
Figure~S4a shows an optical microscope image of the loss test structure, a standard
silicon waveguide at $1550\,\mathrm{nm}$ with a graphite layer ($L = 38\,\mu\mathrm{m}$)
transferred on top.
The silicon waveguide has a width of $w = 450\,\mathrm{nm}$ and a height of
$h = 220\,\mathrm{nm}$; the SiO$_2$ spacer between graphite and waveguide is
$t_0 = 10\,\mathrm{nm}$ thick and the graphite thickness is
$t_{\mathrm{g}} = 12\,\mathrm{nm}$.
Figure~S4b presents the measured transmission data, from which a propagation loss of
$0.91\,\mathrm{dB/\mu m}$ is extracted.

\begin{figure}[htbp]
  \centering
  \includegraphics[width=\textwidth]{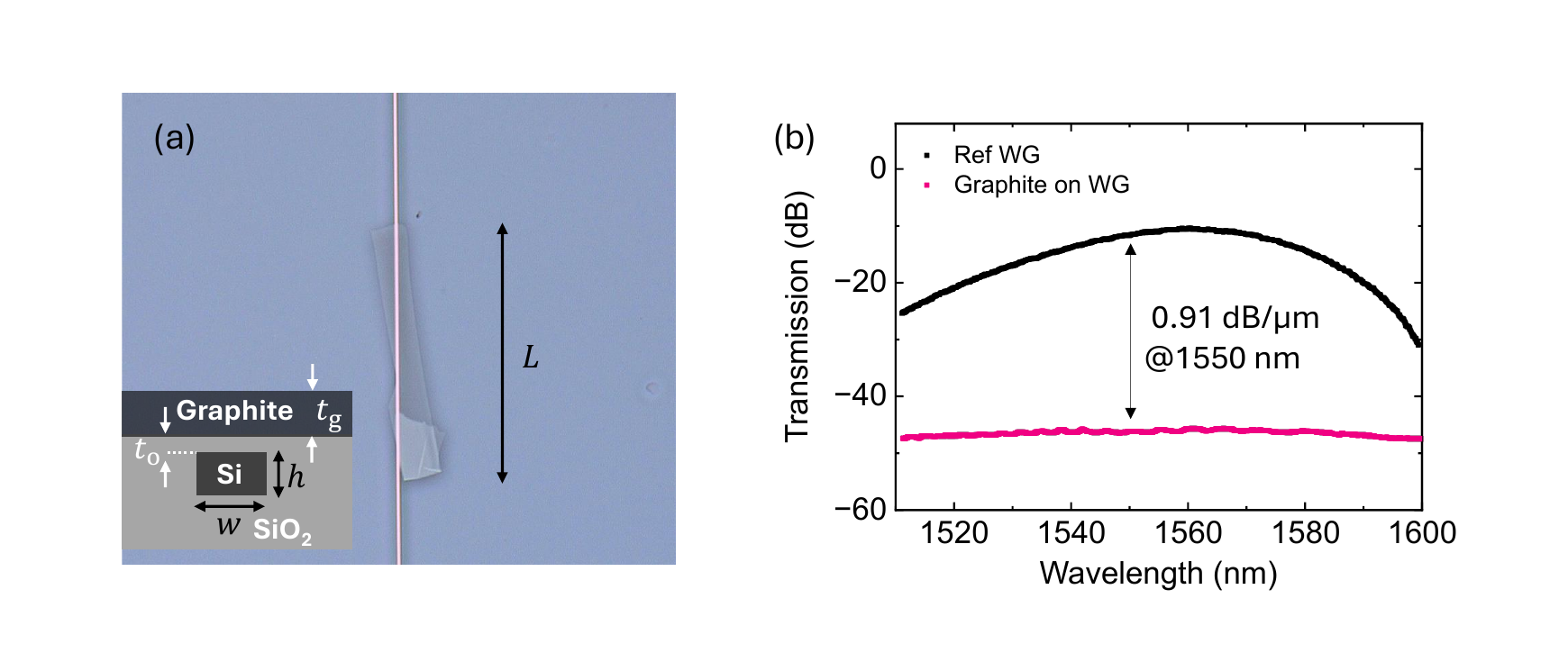}
  \caption{(a)~Microscope image of the loss characterisation waveguide with graphite
           on it and (b)~transmission spectra of the reference waveguide and the
           waveguide with graphite.}
  \label{sfig:4}
\end{figure}

\textit{Absorbed power calculation.}
The graphite crystal used in this work is NGS Flaggy graphite~\cite{NGS_Flaggy_Graphite},
for which the refractive index is not readily available.
To estimate it, we start from handbook values of highly oriented pyrolytic graphite
(HOPG) from HQ Graphene~\cite{SONG20181079}.
At $1550\,\mathrm{nm}$, the handbook values are $n = 2.5276$ and $k = 1.6688$,
yielding a simulated propagation loss of $1.12\,\mathrm{dB/\mu m}$, which exceeds the
measured $0.91\,\mathrm{dB/\mu m}$.
To resolve this discrepancy we tune the Drude damping rate---preserving
Kramers--Kronig relations---from $3.5\times10^{14}\,\mathrm{rad/s}$ to
$2.45\times10^{14}\,\mathrm{rad/s}$, yielding revised constants $n = 2.38$ and
$k = 1.23$ that reproduce the measured loss.
The deviation from handbook values is attributed to differences in crystal quality
between NGS Flaggy graphite and HOPG.

Incorporating these fitted optical constants into simulations of the screened
WSe$_2$--graphene--WSe$_2$ waveguide yields an additional loss of
$0.86\,\mathrm{dB/\mu m}$ from the graphite layer.
The total absorbed power in graphene is obtained by integrating the absorption per
length along the propagation direction (Fig.~S6).
The absorbed power per unit length for the unscreened and screened configurations is
shown in Figs.~S5a and S5b, respectively; in the screened case the graphite layer
starts $7.5\,\mu\mathrm{m}$ into the stack region, causing the absorbed power to drop
much more steeply.
The photovoltages and cooling times in the main text are plotted against these
integrated absorbed powers.

\begin{figure}[htbp]
  \centering
  \includegraphics[width=\textwidth]{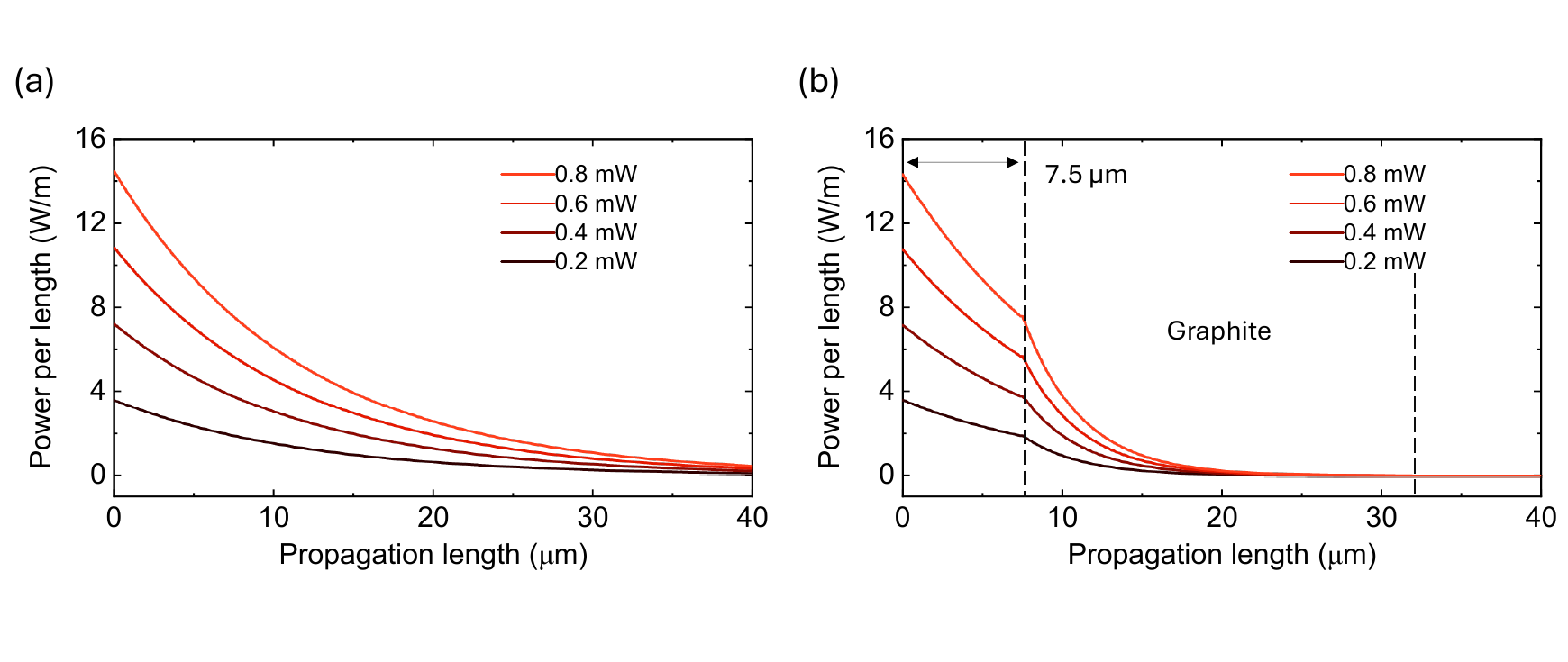}
  \caption{Absorbed power per unit length for the (a)~unscreened and (b)~screened
           cases at different input powers.}
  \label{sfig:5}
\end{figure}

\begin{figure}[htbp]
  \centering
  \includegraphics[width=\textwidth]{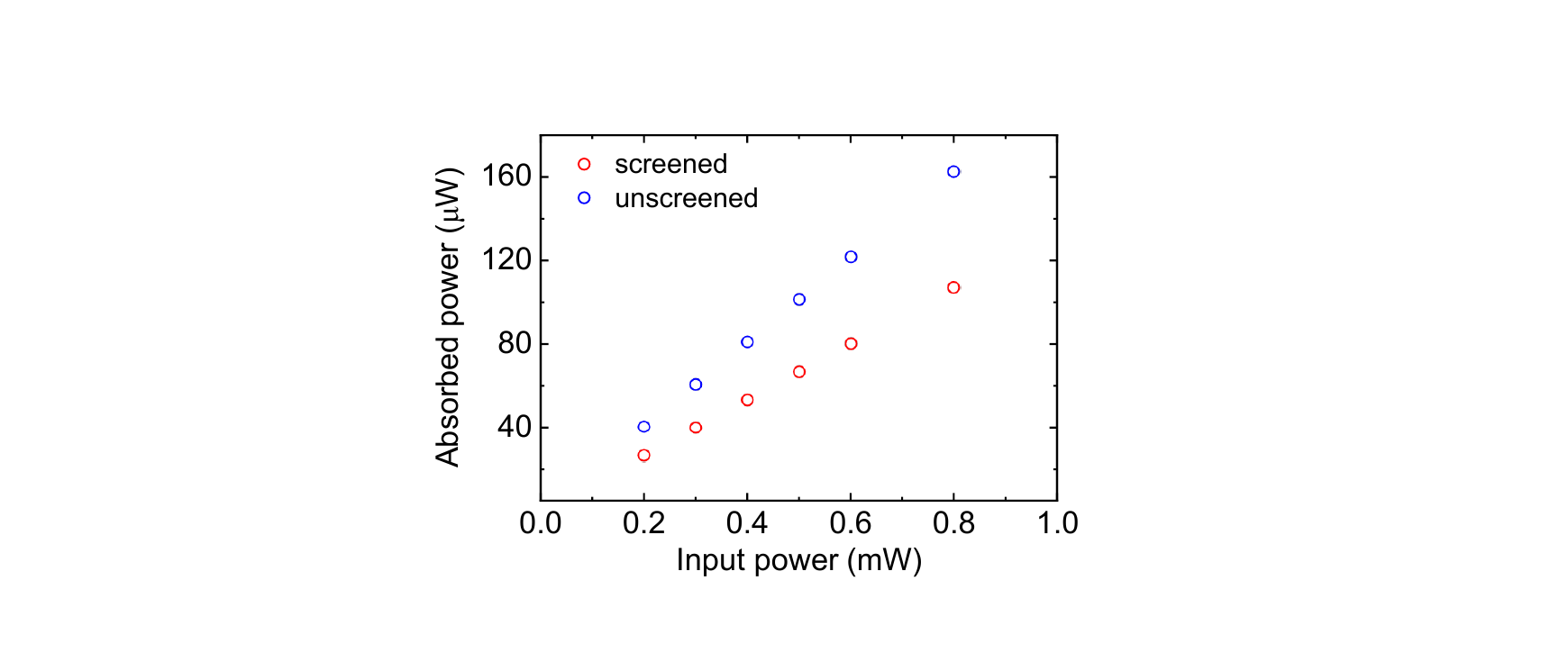}
  \caption{Total graphene absorbed power for different waveguide input powers.}
  \label{sfig:6}
\end{figure}

\section*{Note 4. Simulation to recover the measured photovoltages using cooling time data}

The measured photovoltages under zero bias are attributed to the photothermoelectric
(PTE) effect, which arises from a photon-induced electron temperature gradient
combined with asymmetric doping in the channel.
Under zero-bias conditions, thermally driven carriers accumulate near the contacts,
generating a voltage that enforces zero net current
($\nabla\cdot\mathbf{J} = 0$).
This voltage corresponds to the PTE photovoltage.

\textit{Electron temperature profiles.}
To illustrate the effect of proximity screening on the PTE effect, we performed a
heat transport simulation under CW laser input in steady state.
The 1D heat transport equation is
\begin{equation}
C_e \frac{dT_e}{dt}
=
Q
-
C_e\frac{T_e - T_0}{\tau_{\mathrm{cool}}}
-
K_e \frac{\partial^2 T_e}{\partial x^2}.
\end{equation}
Note that the photovoltage simulations presented below use a 2D framework; the 1D
case is shown here for clarity.
A cross-sectional simulation provides the heating source profile (Fig.~S7a).
Figure~S7b shows that increasing the cooling time from $\tau = 1.2\,\mathrm{ps}$ to
$2.3\,\mathrm{ps}$ raises the maximum electron temperature by approximately
$20\,\mathrm{K}$ at an absorbed power of $60\,\mu\mathrm{W}$.

\begin{figure}[htbp]
  \centering
  \includegraphics[width=\textwidth]{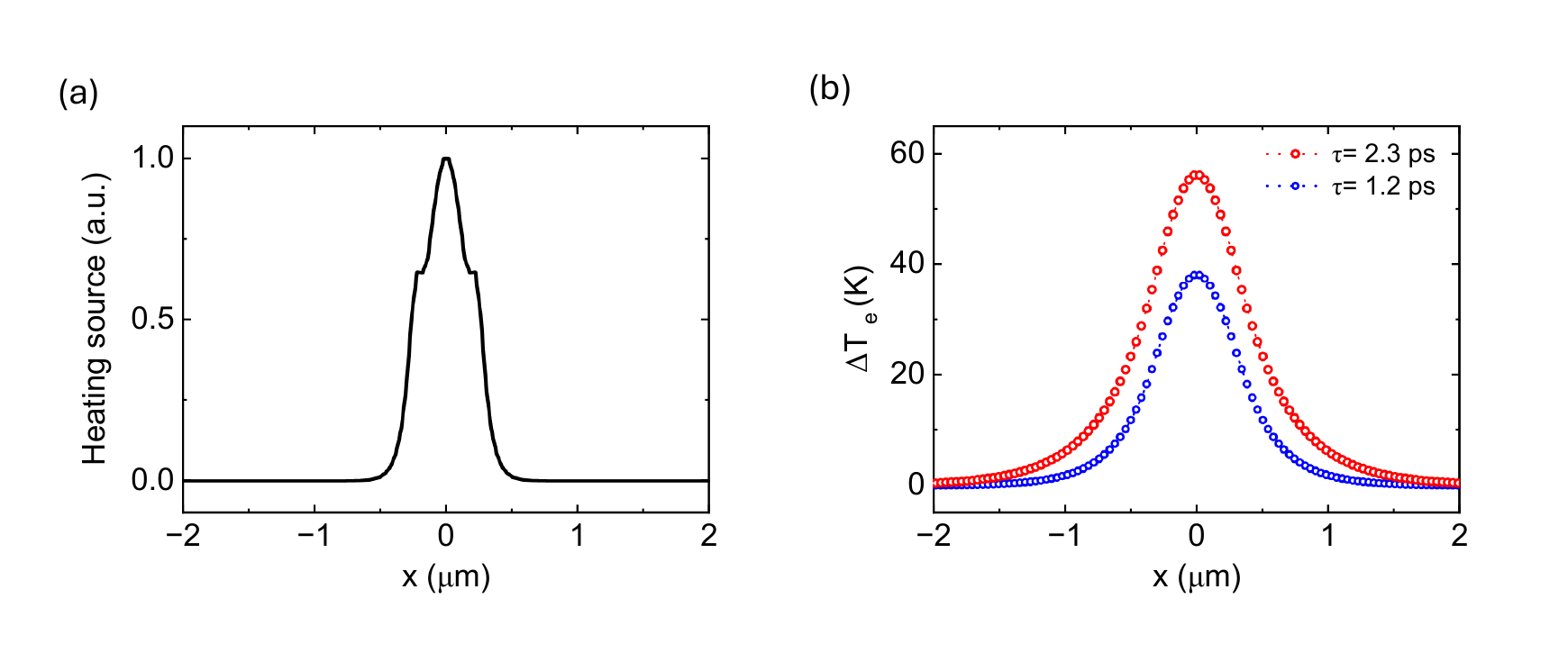}
  \caption{(a)~Heating source and (b)~electron temperature at the input cross section.}
  \label{sfig:7}
\end{figure}

\textit{Modeling of waveguide-coupled photovoltages.}
Due to fabrication limitations the device lacks a well-defined electrostatic p--n
junction.
We therefore assume an effective asymmetric doping of $\pm1.0\,\mathrm{meV}$, which
yields results consistent with the experimental data.
The Dirac-point voltage is $U_{\mathrm{Dirac}} = +140\,\mathrm{mV}$, corresponding
to $E_F = -71.7\,\mathrm{meV}$ at $U_{\mathrm{G1}} = U_{\mathrm{G2}} = 0\,\mathrm{V}$.
The carrier mobility is assumed to be
$10{,}000\,\mathrm{cm^2\,V^{-1}\,s^{-1}}$ based on the Raman characterisation in
Note~1.
The intrinsic doping level is consistent with the measured zero-gate channel
resistance of $\approx400\,\Omega$ for both cases.
The model includes the temperature-dependent Fermi-level shift due to elevated carrier
temperature.

Photovoltage simulations are based on the 1D heat equation for the
propagation-direction-averaged carrier-temperature profile~\cite{Antidormi2021,Massicotte2021}:
\begin{equation}
\frac{\partial T_C(x,t)}{\partial t}
=
\frac{1}{C_e(x,t)}
\frac{\partial}{\partial x}
\!\left[
\kappa_e(x,t)
\frac{\partial T_C(x,t)}{\partial x}
\right]
+
\frac{\dot{Q}_{\mathrm{loss}}(x,t)}{C_e(x,t)}
+
\frac{p_{\mathrm{abs}}(x,t)}{C_e(x,t)},
\end{equation}
where $T_C$ is the carrier temperature, $C_e$ is the electronic heat capacity per
unit area, $\kappa_e$ is the electronic thermal conductivity, $\dot{Q}_{\mathrm{loss}}$
is the carrier-cooling power density, and $p_{\mathrm{abs}}$ is the absorbed optical
power density taken from Fig.~S6.

The electronic heat capacity and thermal conductivity are calculated from graphene's
finite-temperature carrier statistics and Boltzmann transport
integrals~\cite{Antidormi2021,Massicotte2021}.
The thermal conductivity is obtained from the transport moments
\begin{equation}
K_j
=
\int
(E-\mu)^j \Sigma(E)
\left(-\frac{\partial f}{\partial E}\right)dE,
\end{equation}
with
\begin{equation}
\Sigma(E)
=
\frac{e^2 v_F^2 \tau_p(E)D(E)}{2},
\qquad
D(E)=\frac{2|E|}{\pi(\hbar v_F)^2}.
\end{equation}
The momentum relaxation time $\tau_p(E)$ includes acoustic-phonon and
remote-impurity scattering~\cite{Shishir2009}.

Carrier cooling is modeled by a linear heat-loss term,
\begin{equation}
\dot{Q}_{\mathrm{loss}}(x,t)
=
-C_e(x,t)\,\gamma_c
\left[T_C(x,t)-T_L\right],
\qquad
\gamma_c=\frac{1}{\tau_c}.
\end{equation}
In the unscreened case a power-dependent $\tau_c$ is used; in the screened case a
fixed $\tau_c$ is assumed.
The lattice temperature $T_L$ is kept constant.
The heat equation is solved self-consistently with temperature-dependent transport
coefficients using an implicit Crank--Nicolson finite-difference
scheme~\cite{Crank1947}.

\begin{figure}[htbp]
  \centering
  \includegraphics[width=\textwidth]{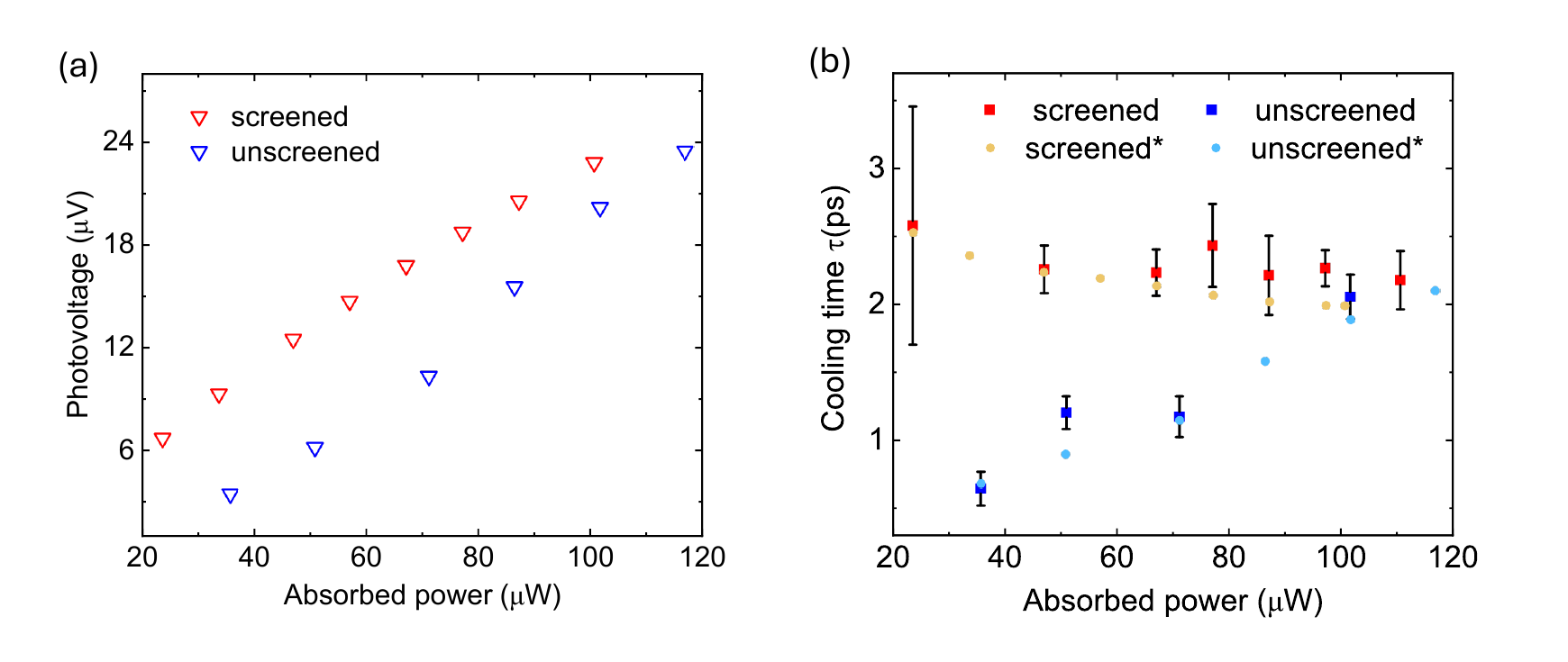}
  \caption{(a)~Simulated photovoltages from the nonlinear photothermoelectric
           heat-transport model using $U_{\mathrm{Dirac}} = +140\,\mathrm{mV}$
           ($E_F = -71.7\,\mathrm{meV}$), asymmetric doping of $\pm1.0\,\mathrm{meV}$,
           and mobility $10{,}000\,\mathrm{cm^2\,V^{-1}\,s^{-1}}$.
           (b)~Cooling times extracted from measured and simulated photomixing
           frequency roll-offs.
           Starred labels (``screened$^*$'', ``unscreened$^*$'') denote cooling times
           extracted from simulated photomixing traces that include heat diffusion,
           after applying the same Lorentzian analysis as in Note~2.}
  \label{sfig:8}
\end{figure}

Figure~S8a shows the simulated photovoltages.
The observed enhancement is reproduced in the intermediate power regime; deviations
at low and high power arise from uncertainties in the effective absorbed power, the
assumed Seebeck asymmetry, and the cooling-time data used as model input.

Figure~S8b compares cooling times from measured and simulated photomixing traces.
In all simulations, Eq.~(S5) is solved in space and time including heat diffusion, and
the resulting photovoltage is analyzed with the same Lorentzian fitting procedure as
in Note~2.
The starred data points represent the cooling times that the photomixing technique
would extract from the simulated device response.
The nonlinear heat-transport model captures the main trends of the measured cooling
times and photovoltage response.

To further clarify the relation between the microscopic input cooling time and the
photomixing-extracted cooling time, we performed an additional simulation with a fixed
$\tau_c = 4\,\mathrm{ps}$, using $U_{\mathrm{G1}} = 0.3\,\mathrm{V}$ and
$U_{\mathrm{G2}} = -0.3\,\mathrm{V}$ to form a well-defined p--n junction, and a
Gaussian optical heating profile with $D_{1/e^2} = 5\,\mu\mathrm{m}$.

\begin{figure}[htbp]
  \centering
  \includegraphics[width=0.52\textwidth]{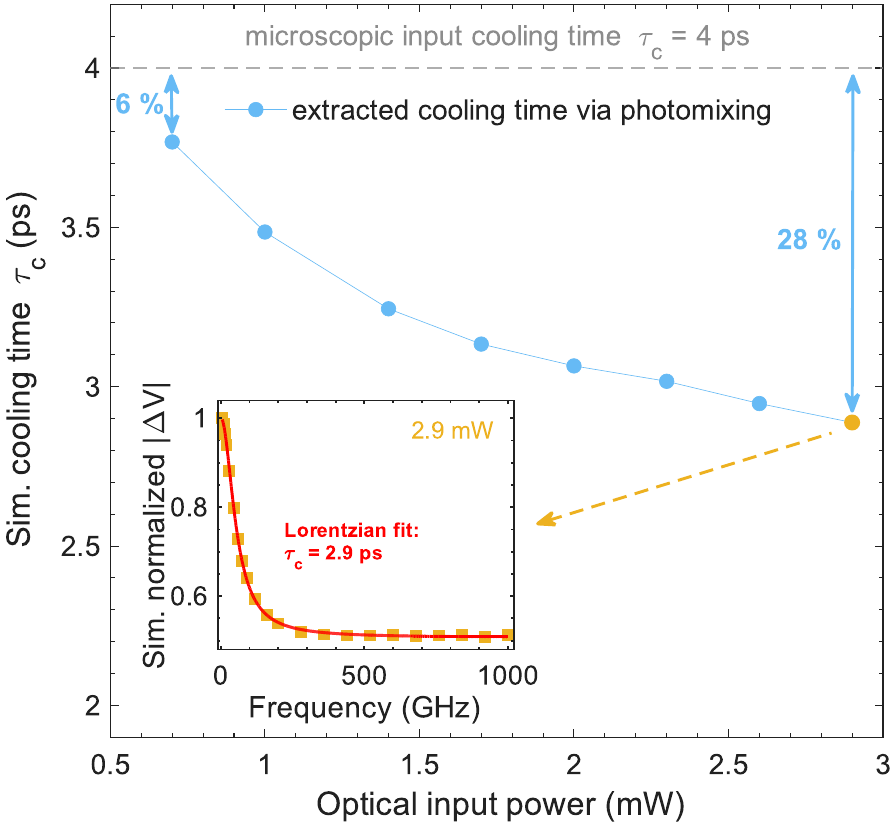}
  \caption{Simulated cooling-time extraction for a fixed microscopic input
           $\tau_c = 4\,\mathrm{ps}$.
           Dashed line marks the input value; percentages indicate the relative
           extraction error.
           Inset: normalized high-power photomixing roll-off with Lorentzian fit.}
  \label{sfig:9}
\end{figure}

As shown in Fig.~S9, the photomixing-extracted cooling time can be systematically
smaller than the microscopic input $\tau_c$, with the underestimation increasing at
higher optical power.
The deviation arises from the spatially extended, nonlinear nature of the
heat-transport problem: lateral heat diffusion, temperature-dependent heat capacity
and thermal conductance, and temperature-dependent chemical potential all modify the
dynamic carrier-temperature profile beyond the approximations of simplified
photomixing analyses~\cite{Jadidi_PRL}.
This also suggests that experimental cooling times from such analyses may underestimate
the true microscopic cooling time.
A broader parameter-space study will be presented in a forthcoming publication.

\section*{Note 5. Cooling time of closer proximity screening}

The Supplementary Material of Ref.~\cite{Wang2026} describes an approach for
examining the cooling dynamics of hot carriers in graphene, arising from the combined
effects of carrier--carrier scattering, optical phonon emission, and
optical-to-acoustic phonon relaxation.
Using that approach, we carried out simulations investigating the impact of dielectric
screening.
As shown in Fig.~S10, the cooling time scales approximately linearly with permittivity.

\begin{figure}[htbp]
  \centering
  \includegraphics[width=0.6\textwidth]{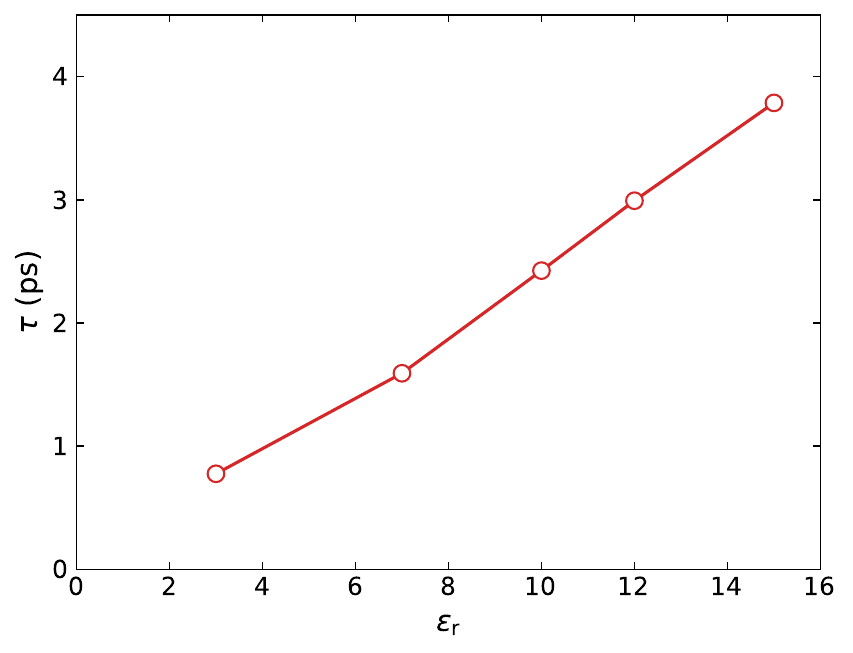}
  \caption{Hot-carrier cooling time as a function of surrounding permittivity,
           calculated using the approach of Ref.~\cite{Wang2026}.}
  \label{sfig:10}
\end{figure}

When placing a metal near the graphene layer, the Coulomb interaction becomes
\begin{equation}
V(q) = \left(1 - \x{e}^{-2qd}\right)\frac{e^2}{2\epsilon q},
\label{eq:coul_metal}
\end{equation}
where $d$ is the graphene--metal distance and $q$ is the carrier momentum.
In the limit of large $d$, the bare Coulomb potential is recovered.
In the limit $qd \ll 1$, the interaction reduces to $V(q) = e^2 d/\epsilon$, so that
$\tau \propto 1/d$, analogous to the $\tau \propto \epsilon$ dependence seen above.

To verify that we operate in the regime $qd \ll 1$, we note that for clean WSe$_2$-encapsulated graphene the dominant cooling mechanism is optical phonon emission,
followed by fast re-thermalization via carrier--carrier
scattering~\cite{doi:10.1021/acsnano.0c10864}.
At low optical powers the carrier distribution is near room temperature, so momentum
transfers during re-thermalization are restricted to
$q \lesssim k_{\mathrm{B}}T/\hbar v_{\mathrm{F}} \approx 1/25\,\mathrm{nm}^{-1}$.
For $d \le 5.5\,\mathrm{nm}$ we have $qd \ll 1$, confirming that $\tau \propto 1/d$
in our device geometry.

\printbibliography[title={References}]
\end{refsection}

\end{document}